\documentclass[aip, reprint, superscriptaddress]{revtex4-2}

\usepackage{amsmath}
\usepackage{amssymb}
\usepackage{xspace}
\usepackage{color}
\usepackage{graphicx}
\usepackage{hyperref}
\usepackage{natbib}

\usepackage{ifpdf}
\ifpdf
\fi

\newcommand{\eq}[1]{(\ref{#1})}
\newcommand{\Eq}[1]{Eq.~(\ref{#1})}

\newcommand{\Fig}[1]{Fig.~\ref{#1}}
\newcommand{\Sec}[1]{Sec.~\ref{#1}}

\newcommand{\ie}{{i.e.,\/}\xspace}

\newcommand{\pd}{\partial}

\newcommand{\mc}[1]{\mathcal{#1}}
\newcommand{\bd}[1]{\boldsymbol{#1}}

\newcommand{\moyalsin}[1]{\{\!\{#1\}\!\}}
\newcommand{\moyalcos}[1]{[[#1]]}
\newcommand{\avg}[1]{\langle{#1}\rangle}

\begin{document}
\title{Wave-kinetic approach to zonal-flow dynamics: recent advances}
\author{Hongxuan Zhu}\altaffiliation{Author to whom correspondence should be addressed: hzhu@pppl.gov}
\affiliation{Princeton Plasma Physics Laboratory, Princeton, NJ 08543}
\author{I.~Y. Dodin}
\affiliation{Princeton Plasma Physics Laboratory, Princeton, NJ 08543}
\affiliation{Department of Astrophysical Sciences, Princeton University, Princeton,
NJ 08544}
\begin{abstract}
Basic physics of drift-wave turbulence and zonal flows has long been studied within the framework of wave-kinetic theory. Recently, this framework has been re-examined from first principles, which has led to more accurate yet still tractable ``improved'' wave-kinetic equations. In particular, these equations reveal an important effect of the zonal-flow ``curvature'' (the second radial derivative of the flow velocity) on dynamics and stability of drift waves and zonal flows. We overview these recent findings and present a consolidated high-level picture of (mostly quasilinear) zonal-flow physics within reduced models of drift-wave turbulence. 
\end{abstract}
\maketitle
\section{Introduction}
It is well known that sheared $\bd{E}\times\bd{B}$ flows, including equilibrium flows and spontaneously generated zonal flows (ZFs), can reduce the level of drift-wave (DW) turbulence in fusion plasmas and play a crucial role in the transition between regimes with low and high (L--H) confinement. (For reviews, see, e.g., Refs.~\onlinecite{Diamond05,Fujisawa08,Connaughton15,Burrell20}.)  Numerical simulations  have also shown that ZFs can even completely suppress turbulence near the instability threshold,  which effect is known as the Dimits shift.\cite{Hammett93,Dimits96,Lin98,Dimits00,Rogers00,Mikkelsen08,Kobayashi12} Because of this, DW--ZF interactions have been attracting much attention over the last decades and  studied extensively.

One of the theoretical frameworks used for studying DW--ZF interactions is the wave-kinetic theory of inhomogeneous DW turbulence.\cite{Diamond05,Malkov01,Malkov01PP,Smolyakov99,Kaw01,Mendoncca03,Trines05,Wang09,Kosuga18,Sasaki18,Dyachenko92,Manin94,Balk90a,Balk90b} Within this framework, which assumes ZFs to have scales much larger than the characteristic DW wavelength, DWs are described as effective classical particles, sometimes called ``driftons''.\cite{Mendoncca03} The drifton phase-space density is described by the wave-kinetic equation (WKE). The ZF velocity enters the WKE through the drifton Hamiltonian and serves as a collective field through which driftons interact. This approach has been fairly successful; for example, it has yielded predator--prey models that explain some aspects of the  L--H transition.\cite{Diamond94,Malkov01,Kim03} However, because the ``traditional'' WKE relies on \textit{ad~hoc} assumptions and is  not entirely rigorous, the potential of the wave-kinetic approach in application to DW turbulence is yet to be fully appreciated.

Recently, the wave-kinetic description of inhomogeneous DW turbulence has been re-examined from first principles and applied to make quantitative predictions in a number of problems.\cite{Parker16,Ruiz16,Ruiz19,Zhu18a,Zhu18b,Zhu18c,Zhu19,Zhu20JPP,Zhu20PRL,Zhou19,Zhou20,Tsiolis20} Although still limited to simplified DW models, those results indicate that important qualitative physics has been overlooked in the past but can be described transparently if the wave-kinetic formalism is properly amended. In particular, the  ``improved'' WKE reveals an important effect of the ZF ``curvature'' (the second radial derivative of the zonal velocity) on dynamics and stability of DWs and ZFs. Here, we overview these results and present a consolidated high-level picture of DW--ZF interactions and ZF stability. Our analysis is mostly based on the modified Hasegawa--Mima model\cite{Hasegawa77,Hasegawa78,Krommes00} and on the quasilinear approximation (which neglects DW--DW interactions but keeps DW--ZF coupling), although more general models are also considered. A special focus is made on understanding drifton phase-space dynamics, associated solitary structures, merging and splitting of ZFs, as well as the Kelvin--Helmholtz instability (KHI) and the tertiary instability. Notably, while the KHI and the tertiary instability are often confused with each other, they have very different properties, as discussed below. We also briefly mention the connection between these findings and the recent progress in analytic understanding of the Dimits shift.\cite{St-Onge17b,Zhu20JPP,Zhu20PRL,Ivanov20,Qi20,Hallenbert20}

This paper is organized as follows. Basic model equations of DWs are introduced in \Sec{sec:models}. The wave-kinetic framework and the resulting traditional WKE is outlined in \Sec{sec:TWKE}. The improved WKE derived from first principle is introduced in \Sec{sec:IWKE}. Several recent discoveries based on the improved WKE, along with some new findings, are reviewed in \Sec{sec:dynamics}. The tertiary instability and the Dimits shift are discussed in \Sec{sec:curvature}. A summary is presented in \Sec{sec:summary}.
\section{Basic equations}
\label{sec:models}
For the most part of this paper, we shall use the modified Hasegawa--Mima equation (mHME) as our base model, as it is particularly simple yet captures many features of DW--ZF interactions. \cite{Hasegawa77,Hasegawa78,Krommes00} The mHME describes electrostatic potential fluctuations $\varphi$ in a two-dimensional slab $\bd{x}=(x,y)$ and a uniform magnetic field $\bd{B}=B\hat{\bd{z}}$. Here, $x$ and $y$ correspond to the radial and poloidal directions in tokamaks, respectively.  Ions are assumed cold, while electrons have finite temperature $T_e$ and are assumed to respond adiabatically to $\varphi$ (but also see below). The plasma is assumed to have an equilibrium density profile  $n_0(x)$ parameterized by a constant $\kappa\doteq a/L_n$, where $a$ is the system length and $L_n\doteq(-n_0'/n_0)^{-1}$. (We use $\doteq$ to denote definitions and prime to denote $\pd_x$.)    We normalize time by $a/C_s$ where $C_s\doteq\sqrt{T_e/m_i}$ and $m_i$ is the ion mass,  length by $\rho_s\doteq C_s/\Omega_i$ where $\Omega_i$ is the ion gyrofrequency, and  $\varphi$ by $T_e\rho_s/ea$.  Then, the mHME is written as
\begin{equation}
\pd_t w +\bd{v}\cdot\nabla w-\kappa\pd_y\varphi=f-d.\label{eq:models_mHME}
\end{equation}
Here, $\bd{v}\doteq\hat{\bd{z}}\times\nabla\varphi$ is the $\bd{E}\times\bd{B}$ velocity, $w\doteq\nabla^2\varphi-\tilde{\varphi}$ is minus the ion guiding-center density, and $\nabla^2\doteq\pd_x^2+\pd_y^2$. Note that $w$ is modified compared to the original HME;\cite{Hasegawa77} namely, only the fluctuation part of the potential $\tilde{\varphi}\doteq\varphi-\avg{\varphi}$ contributes to the second term in $w$, where
\begin{equation}
\avg{\varphi}\doteq\frac{1}{L_y}\int_0^{L_y}dy\, \varphi
\end{equation}
is the zonal-averaged part of $\varphi$ and $L_y$ is the domain length in $y$.  This modification is due to the special electron response to the zonal potential.\cite{Dorland93,Hammett93} Also, $f$ and $d$ represent forcing and dissipation, respectively.  

Absent forcing and dissipation, the mHME conserves the total energy $\mc{E}$ and enstrophy $\mc{Z}$, which are defined as
\begin{equation}
\mc{E}\doteq-\frac{1}{2}\int d\bd{x}\,w\varphi,\quad\mc{Z}\doteq\frac{1}{2}\int d\bd{x}\,w^2.\label{eq:models_EZ}
\end{equation}
Also, in this case, a Fourier eigenmode $\varphi\propto e^{i\bd{k}\cdot\bd{x}-i\Omega_{\bd{k}}t}$ is a solution of \Eq{eq:models_mHME}, with the following dispersion relation:
\begin{equation}
\Omega_{\bd{k}}={\kappa k_y}/\bar{k}^2,\quad \bar{k}^2\doteq 1+k_x^2+k_y^2.\label{eq:models_dispersion}
\end{equation}
Because of the assumed electron adiabatic response, $\Omega_{\bd{k}}$ is real, so DWs neither grow nor decay in the mHME model. Thus,  DW turbulence is introduced  through either $f$ or initial conditions. But the situation changes  if the electron response is made nonadiabatic. In particular, the modified Terry--Horton equation (mTHE) \cite{Terry82,Terry83,St-Onge17b,Zhu20JPP} assumes
\begin{equation}
n_e=(1-i\hat{\delta})\tilde{\varphi}.\label{eq:models_TH}
\end{equation}
Here, $n_e$ is the electron density fluctuation normalized by  $n_0\rho_s/a$, and the operator $i\hat{\delta}$ introduces a phase difference between $n_e$ and $\tilde{\varphi}$. Correspondingly, the mTHE has the same form as \Eq{eq:models_mHME} but with
\begin{equation}
w=\nabla^2\varphi-(1-i\hat{\delta})\tilde{\varphi}.
\end{equation}
Assuming $\hat{\delta}e^{i\bd{k}\cdot\bd{x}}=\delta_{\bd{k}}e^{i\bd{k}\cdot\bd{x}}$, the DW frequencies $\Omega_{\bd{k}}$ have the same form as in \Eq{eq:models_dispersion}, but with
\begin{equation}
\bar{k}^2\doteq 1+k_x^2+k_y^2-i\delta_{\bd{k}}.
\end{equation}
This allows for DW instabilities (${\rm Im}\,\Omega_{\bd{k}}>0$), which are called 	``primary'' instabilities.  

Various forms of $\hat{\delta}$ can be used to model different mechanisms of primary instabilities. \cite{Tang78,Terry83,St-Onge17b} Alternatively, $n_e$ can be modeled as a dynamic field. For example, if $n_e$ and $\varphi$ are connected via Ohm's law, one is led to a set of two-field equations known as the modified Hasegawa--Wakatani equation.\cite{Hasegawa83,Wakatani84,Numata07,Zhang20a,Zhang20b} This model can be applied to resistive DWs at the tokamak edge. Similar two-field models have also been proposed for core plasmas, including ones that describe the ion-temperature-gradient (ITG) mode,\cite{Rogers00,Ottaviani97,St-Onge17a,Ivanov20} trapped-electron modes,\cite{Leconte19} etc. However, in this paper, we mostly focus on the mHME.
\section{The traditional wave-kinetic equation}
\label{sec:TWKE}
By applying zonal average to \Eq{eq:models_mHME} and assuming the forcing has $\avg{f}=0$, one can write down separate equations for fluctuations (DWs) and ZFs:
\begin{subequations}
\begin{gather}
\pd_t\tilde{w}+U\pd_y\tilde{w}-(\kappa+U'')\pd_y\tilde{\varphi}=-f_{\rm NL}+f-\tilde{d},\label{eq:models_DW}\\
\pd_tU+\pd_x\avg{\tilde{v}_x\tilde{v}_y}=-\int dx\avg{d}.\label{eq:models_ZF}
\end{gather}
\end{subequations}
Here, $U(x,t)\doteq\pd_x\avg{\varphi}$ is the ZF velocity and 
$f_{\rm NL}\doteq\tilde{\bd{v}}\cdot\nabla\tilde{w}-\avg{\tilde{\bd{v}}\cdot\nabla\tilde{w}}$ represents nonlinear interactions of DWs. These interactions are often neglected (\ie $f_{\rm NL}$ is replaced with zero), which is known as the quasilinear approximation.\cite{Parker13,Parker14} Importantly, the total energy and enstrophy [\Eq{eq:models_EZ}] are still conserved in this case.

From here, we briefly review the existing wave-kinetic theory as follows. Let us assume the quasilinear approximation and
\begin{equation}
\epsilon\doteq\max(\lambda_{\rm DW},\rho_s)/\lambda_{\rm ZF}\ll 1,\label{eq:WKE_GO}
\end{equation}
where $\lambda_{\rm DW}$ and $\lambda_{\rm ZF}$ are the typical spatial scales of DWs and ZFs, respectively. This condition is often satisfied in simulations,\footnote{However, as mentioned in Ref.~\onlinecite{Parker16}, experimental results in Refs~\onlinecite{Fujisawa08,Gupta06,Hillesheim16} suggest that the ZF wavelength can as well be comparable to those of DWs. Then, the Wigner--Moyal equation outlined in \Sec{sec:IWKE} should be used.} and $\epsilon\to 0$ is known as the geometrical-optics (GO) limit.  Within this limit, DWs are considered as quasimonochromatic packets of the form
\begin{gather}
\tilde{\varphi}(\bd{x}, t) = {\rm Re}[\varphi_a(\bd{x}, t) e^{i \theta_a}].
\end{gather}
Here, $a$ is the packet index, and the complex envelope $\varphi_a$ is assumed to vary slowly compared to the real phase $\theta_a$. The gradient of this phase, $\bd{k}_a \doteq \nabla \theta_a$, serves as the local wavevector, and $\Omega_a \doteq - \partial_t \theta_a$ is the local frequency, which satisfies the dispersion relation $\Omega_a = \bar{\Omega}_{\bd{k}}(\bd{x}, t)$. Note that $\bar{\Omega}_{\bd{k}}$ differs from $\Omega_{\bd{k}}$ [\Eq{eq:models_dispersion}], and the difference between the two is the Dopper shift due to the ZF velocity:
\begin{equation}
\bar{\Omega}_{\bd{k}}=\Omega_{\bd{k}}+k_yU.\label{eq:TWKE_Omegabar}
\end{equation}
For simplicity (and at the expense of rigor), we henceforth omit the index $a$ and label envelopes with $\bd{k}$ instead.  Envelopes propagate according to ray equations:\cite{Dyachenko92,Tracy14book}
\begin{equation}
\frac{d\bd{x}}{dt}=\frac{\pd\bar{\Omega}_{\bd{k}}}{\pd\bd{k}},\quad\frac{d\bd{k}}{dt}=-\frac{\pd\bar{\Omega}_{\bd{k}}}{\pd\bd{x}}.\label{eq:TWKE_ray}
\end{equation}
Because $\pd\bar{\Omega}_{\bd{k}}/\pd y=0$, one has $dk_y/dt=0$; \ie  $k_y$ is conserved and can be considered as a parameter.  

From \Eq{eq:models_EZ}, the energy density of a DW packet in homogeneous plasmas is (up to a constant factor)
\begin{equation}
\mc{E}_{\bd{k}}=(1+k^2)|\varphi_{\bd{k}}|^2.
\end{equation}
Then,  the wave action density is
\begin{equation}
n_{\bd{k}}\doteq\frac{\mc{E}_{\bd k}}{\Omega_{\bd{k}}}=\frac{(1+k^2)^2}{\kappa k_y}|\varphi_{\bd{k}}|^2,\label{eq:TWKE_nk}
\end{equation}
and the total wave action $\mc{I}\doteq\int n_{\bd{k}}\,d\bd{x}$ is conserved.\cite{Dyachenko92,Tracy14book} In the presence of ZFs, $\mc{I}$ is still conserved in the GO limit.\cite{Mattor94,Brizard96} Hence, a DW envelope can be considered as an effective particle, a drifton. Let us define the drifton density $N(\bd{x},\bd{k},t)$ in phase space $(\bd{x}, \bd{k})$ as the sum of $(1+k^2)^2 |\varphi_{\bd{k}}|^2$ over all envelopes. (An explicit definition can be found in Ref.~\onlinecite{Smolyakov99}.)  Note that the factor $\kappa k_y$ is omitted since $\kappa$ is a constant and $k_y$ is conserved.  Then, by analogy with the Liouville equation for classical particles, one  obtains what we call the traditional WKE:
\begin{equation}
\pd_t N=\lbrace \bar{\Omega}_{\bd{k}},N\rbrace,\label{eq:WKE_N0}
\end{equation}
where $\bar{\Omega}_{\bd{k}}$ is given by \Eq{eq:TWKE_Omegabar}, or more explicitly,
\begin{equation}
\bar{\Omega}_{\bd{k}}=\kappa k_y/\bar{k}^2+k_y U.
\end{equation} 
Here, forcing and dissipation are neglected and we have introduced the canonical Poisson bracket:
\begin{equation}
\lbrace A,B\rbrace=\frac{\pd A}{\pd\bd{x}}\cdot\frac{\pd B}{\pd{\bd{k}}}-\frac{\pd A}{\pd {\bd{k}}}\cdot\frac{\pd B}{\pd\bd{x}}.\label{eq:TWKE_U}
\end{equation}
Also, \Eq{eq:models_ZF} can be rewritten as\cite{Dyachenko92,Smolyakov00b,Malkov01,Trines05}
\begin{equation}
\frac{\pd  U(x,t)}{\pd t}=\frac{\pd}{\pd x}\int\frac{d\bd{k}}{(2\pi)^2}\frac{\kappa k_xk_y}{\bar{k}^4}\avg{N},\label{eq:WKE_U}
\end{equation}
where ZF damping term $\avg{d}$ has been neglected for simplicity.  (For a more rigorous derivation, see \Sec{sec:IWKE}).

Equations  \eqref{eq:WKE_N0} and \eqref{eq:WKE_U}, which we call the traditional wave-kinetic model, play the same role as the Vlasov--Poisson system in electron plasmas. This means that DW turbulence within the wave-kinetic approach behaves like an effective plasma made of driftons, and the ZF velocity $U$ in $\bar{\Omega}_{\bd{k}}$ serves as a collective field through which driftons interact.  In particular, this helps construct DW--ZF equilibria in analogy with the Bernstein--Greene--Kruskal modes\cite{Bernstein57,Kaw01,Singh14} and explain spatially inhomogeneous distribution of DWs in ZFs.\cite{Sasaki18}
Following the analogy with the Vlasov--Poisson system, one can describe linear collective waves in drifton plasma (i.e., low-amplitude modulational waves in DW turbulence) much like Langmuir waves and derive their dispersion relation accordingly. Assuming a background homogeneous distributions of driftons $N=N_0(\bd{k})$ and a small ZF velocity of the form $U\propto e^{iqx-i\omega t}$, one finds that \cite{Smolyakov00a,Smolyakov00b}
\begin{equation}
\omega=\int\frac{d\bd{k}}{(2\pi)^2}\frac{\kappa k_xk_y^2}{\bar{k}^4}\frac{q^2}{\omega-qv_g}\frac{\pd N_0}{\pd k_x},\label{eq:WKE_dispersion1}
\end{equation}
where $v_g\doteq -2\kappa k_xk_y/\bar{k}^4$
is the group velocity of DWs in the $x$ direction. Equation \eqref{eq:WKE_dispersion1} predicts that, for a broad class of $N_0$, the ZF velocity grows exponentially (${\rm Im}\,\omega > 0$). This is known as the zonostrophic instability,\cite{Farrell07,Srinivasan12,Parker13,Parker14} which is sometimes also called the secondary instability.\cite{Rogers00,Diamond01} The well-known modulational instability of DWs is a special situation of the zonostrophic instability that corresponds to monochromatic DWs, \ie delta-shaped $N_0$.\cite{Manin94,Champeaux01,Connaughton10,Zhu19} 

One issue with the dispersion relation \eqref{eq:WKE_dispersion1} is that  the growth rate typically scales as ${\rm Im}\,\omega\propto q$ and has no upper bound unless viscosity is introduced. This ``ultraviolet divergence'' was first discussed in Ref.~\onlinecite{Parker16}. It originates from the fact that \Eq{eq:WKE_N0}  conserves the DW enstrophy alone but not the total enstrophy, which includes contribution from DWs \emph{and} ZFs (see also \Sec{sec:IWKE}).

Because the traditional WKE \eqref{eq:WKE_N0}  is based on the mHME,  it does not support primary instabilities. Therefore,  forcing  and dissipation are often included, and the traditional WKE is usually considered in the following modified form:
\begin{equation}
\pd_t N=\lbrace \bar{\Omega}_{\bd{k}},N\rbrace+F-D-\Delta\omega N^2,\label{eq:WKE_N}
\end{equation}
where $F$ and $D$ correspond to $f$ and $d$ in \Eq{eq:models_mHME}, and $\Delta\omega N^2$ is an \textit{ad~hoc} term added to model nonlinear DW--DW interactions governed by $f_{\rm NL}$. This equation is commonly used in literature to derive predator--prey models within the homogeneous-turbulence approximation, in analogy with the theory of nonlinear Langmuir waves governed by the Vlasov equation.\cite{Stix92book}  However, the applicability of the wave-kinetic theory is not restricted to homogeneous turbulence.  Below, we show how the WKE can be used fruitfully to study inhomogeneous DWs in ZFs. For that, a more rigorous version of the WKE will be needed, which we present in the next section.

\section{The improved wave-kinetic equation}
\label{sec:IWKE}
An improved WKE can be derived from first principles following Refs.~\onlinecite{Parker16,Ruiz16}. For simplicity, let us again neglect forcing ($f = 0$) and dissipation ($d=0$) and adopt the quasilinear approximation ($f_{\rm NL} = 0$). Then, \Eq{eq:models_DW} can be written as a Schr\"odinger equation
\begin{equation}
i\pd_t \tilde{w}=\hat{H}\tilde{w},\label{eq:IWKE_DW}
\end{equation}
where the Hamiltonian $\hat{H}$ is given by
\begin{equation}
\hat{H}\doteq U\hat{k}_y+(\kappa+U'')\hat{k}_y\hat{\bar{k}}^{-2},\label{eq:IWKE_DWH}
\end{equation}
with notations $\hat{k}_y\doteq-i\pd_y$ and $\hat{\bar{k}}^2\doteq 1-\nabla^2$. Let us also introduce the so-called Wigner function, which is the Fourier spectrum of a two-point correlation function of the turbulent field $\tilde{w}$:\cite{Wigner32}
\begin{equation*}
W(\bd{x},\bd{k},t)\doteq\int d\bd{s}\,e^{-i\bd{k}\cdot\bd{s}}\tilde{w}(\bd{x}+\bd{s}/2,t)\tilde{w}(\bd{x}-\bd{s}/2,t).\label{eq:WKE_Wigner}
\end{equation*}
Then, \Eq{eq:IWKE_DW}  leads to what is known as the Wigner--Moyal equation for $W$:\footnote{A similar approach can be used also in more complex turbulence models, including those where turbulence is described by multiples fields. In that case, $W$ is generally a matrix. See Ref.~\onlinecite{Tsiolis20}.} 
\begin{equation}
\pd_t W=\moyalsin{H_H,W}+\moyalcos{H_A,W},\label{eq:IWKE_WME}
\end{equation}
where $H_H$ and $H_A$  are the Weyl symbols of the Hermitian and anti-Hermitian parts of $\hat{H}$:
\begin{subequations}
\begin{gather}
H_H=\kappa k_y/\bar{k}^2+k_yU+\moyalcos{U'',k_y/\bar{k}^2}/2,\\
H_A=\moyalsin{U'',k_y/\bar{k}^2}/2.
\end{gather}
\end{subequations}
(Recall that the prime denotes $\pd_x$.) Here, the Moyal brackets are defined as\cite{Moyal49,Groenewold46} 
\begin{equation*}
\lbrace\!\lbrace A,B \rbrace\!\rbrace\doteq -i(A\star B - B\star A),\quad [[A,B]]\doteq A\star B + B\star A.
\end{equation*}
The symbol $\star$ is the Moyal star product:
\begin{equation}
 A\star B\doteq A e^{i\hat{\mc{L}}/2} B,\quad \hat{\mc{L}}\doteq \frac{\overleftarrow{\pd}}{\pd \bd{x}}\cdot\frac{\overrightarrow{\pd}}{\pd \bd{k}}-\frac{\overleftarrow{\pd}}{\pd \bd{k}}\cdot\frac{\overrightarrow{\pd}}{\pd \bd{x}},\label{eq:IWKE_star}
\end{equation}
where the overhead arrows in $\hat{\mc{L}}$ indicate the directions in which the derivatives act on. (For details, see Refs.~\onlinecite{Ruiz16,Mendoncca11}.) Direct numerical solutions of the Wigner--Moyal equation have been reported in Refs.~\onlinecite{Zhu18c,Zhou19}. Numerical solutions of steady-state DW--ZF equilibria have also been reported in Refs.~\onlinecite{Parker13,Parker14} using the mathematically equivalent second-order cumulant expansion theory.

The Wigner--Moyal equation treats driftons as \textit{quantumlike} particles in that they can have finite wavelength. The WKE is then obtained as a rigorous reduction of the Wigner--Moyal equation in the GO limit [\Eq{eq:WKE_GO}]. In this limit, the Moyal brackets are simplified as
\begin{equation}
\moyalsin{A,B}\approx\lbrace A,B\rbrace,\quad
\moyalcos{A,B}\approx 2AB.
\end{equation}
Consequently, \Eq{eq:IWKE_WME}  becomes what we call the improved WKE:\cite{Parker16,Ruiz16} 
\begin{equation}
\pd_t W=\lbrace\mc{H},W\rbrace+2\Gamma W,\label{eq:IWKE_W}
\end{equation}
with the following Hamiltonian:
\begin{subequations}
\label{eq:IWKE_H}
\begin{gather}
\mc{H}\doteq(\kappa+U'')k_y/\bar{k}^2+k_y U,\\
\Gamma\doteq -k_xk_yU'''/\bar{k}^4.
\end{gather}
\end{subequations}
Meanwhile, the evolution of $U$ is governed by\cite{Parker16,Ruiz16}
\begin{equation}
\frac{\pd  U(x,t)}{\pd t}=\frac{\pd}{\pd x}\int\frac{d\bd{k}}{(2\pi)^2}\frac{\kappa k_xk_y}{\bar{k}^4}\avg{W},\label{eq:IWKE_U}
\end{equation}
which is the same as \Eq{eq:WKE_U} but with $N$ replaced by $W$.

Equations \eqref{eq:IWKE_W} and \eqref{eq:IWKE_U} are the improved wave-kinetic model.
Unlike the traditional WKE \eqref{eq:WKE_N0}, it retains $U''$  and $U'''$. The nonzero $\Gamma$  implies that the number of driftons (DW enstrophy) is not truly conserved. Instead, the WKE has the same conservation laws as the  mHME, \ie conservation of the total energy $\mc{E} = \mc{E}_{\rm DW} + \mc{E}_{\rm ZF}$ and enstrophy  $\mc{Z} = \mc{Z}_{\rm DW} + \mc{Z}_{\rm ZF}$,\cite{Ruiz16} where
\begin{subequations}
\begin{gather}
\mc{E}_{\rm DW}=\frac{1}{2}\int\frac{d\bd{k}\,d\bd{x}}{(2\pi)^2}\,\frac{W}{\bar{k}^2},\quad\mc{E}_{\rm ZF}=\frac{1}{2}\int d\bd{x}\, U^2,\\
\mc{Z}_{\rm DW}=\frac{1}{2}\int\frac{d\bd{k}\,d\bd{x}}{(2\pi)^2}\, W,\quad\mc{Z}_{\rm ZF}=\frac{1}{2}\int d\bd{x}\, (U')^2.
\end{gather}
\end{subequations}

Like the traditional WKE, the improved WKE  can be used to derive the dispersion relation of the linear zonostrophic instability:
\begin{equation}
\omega=\int\frac{d\bd{k}}{(2\pi)^2}\frac{\kappa k_xk_y^2}{\bar{k}^4}\frac{q^2}{\omega-qv_g}\frac{\pd}{\pd k_x}\left[\left(1-\frac{q^2}{\bar{k}^2}\right)W_0\right].\label{eq:IWKE_dispersion}
\end{equation}
As first discussed in Ref.~\onlinecite{Parker16}, the term $q^2/\bar{k}^2$ in \Eq{eq:IWKE_dispersion}, which ensures total enstrophy conservation, also provides a  cutoff of ${\rm Im}\,\omega$ at large $q$.  This prevents the ultraviolet divergence mentioned in \Sec{sec:TWKE}.

Another recent development over the traditional WKE is the replacement of the \textit{ad~hoc} term $-\Delta \omega N^2$ in \Eq{eq:WKE_N} with a rigorous wave--wave collision operator. Based on the weak-turbulence assumption and the quasinormal closure, this collision operator was derived in Ref.~\onlinecite{Ruiz19} using the Wigner--Moyal approach and is given by
\begin{equation}
C[W,W]\doteq S_{\rm NL}[W,W]-2\Gamma_{\rm NL}[W]W.
\end{equation}
(Here, the nonlinear source term $S_{\rm NL}$ is a quadratic function of $W$ and the nonlinear damping rate $\Gamma_{\rm NL}$ depends linearly on $W$.) This collision operator can be placed at the right-hand side of \Eq{eq:IWKE_W} and conserves energy and enstrophy of DWs. A preliminary numerical investigation of this  operator is reported in Ref.~\onlinecite{Ruiz20}.
\section{Linear and nonlinear dynamics from the improved wave-kinetic equation}
\label{sec:dynamics}
\begin{figure}
\includegraphics[width=1\columnwidth]{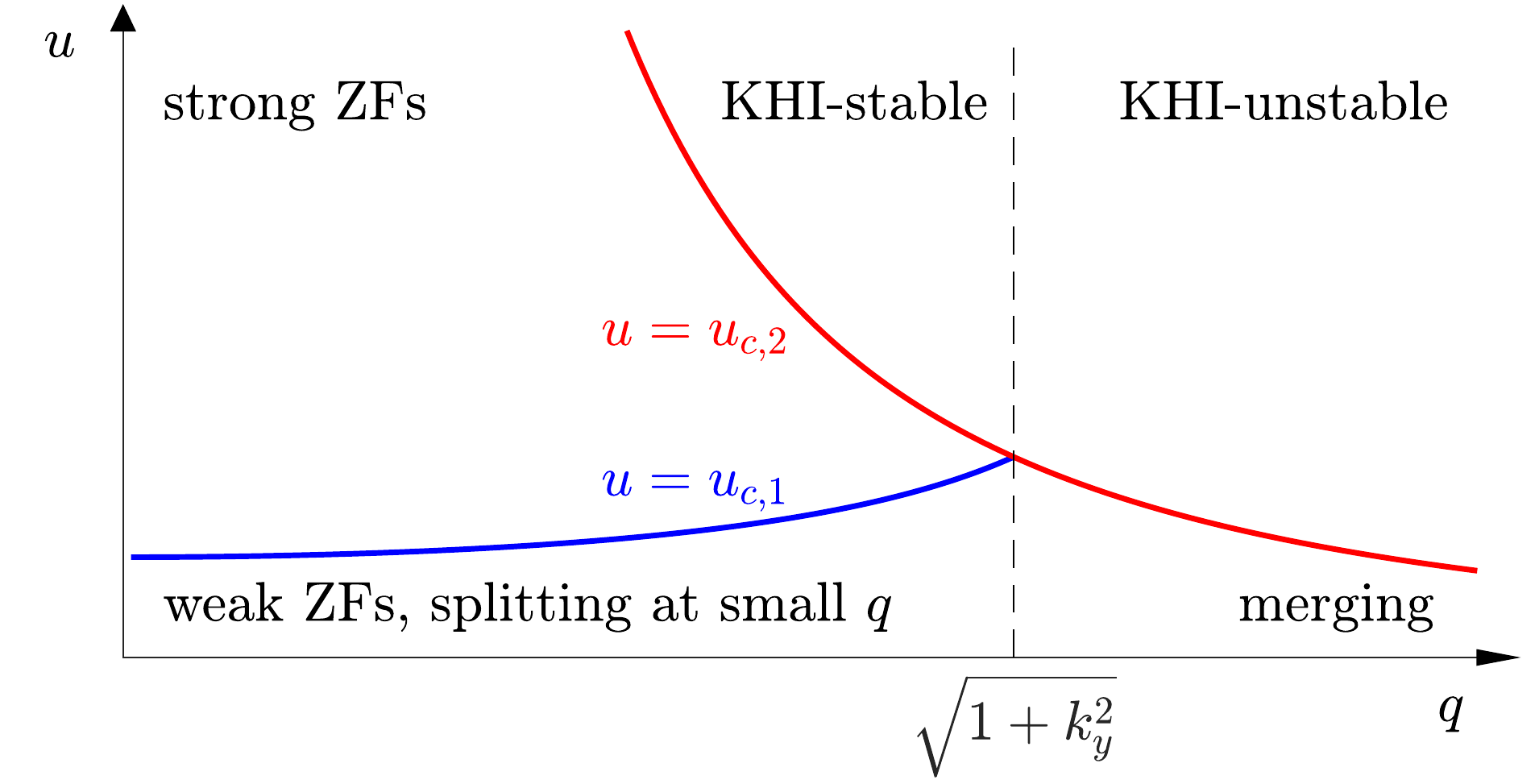}\caption{A  qualitative phase diagram of ZFs summarizing the main results of \Sec{sec:dynamics}. Here, $u$ is the ZF velocity amplitude, $q$ is the ZF wavenumber, and $u_{c,1}$ and $u_{c,2}$ are given by \Eq{eq:traj_uc}. The transition from weak to strong ZFs is discussed in \Sec{subsec:traj}, the KHI is discussed in \Sec{subsec:KHI}, and the splitting and merging of ZFs are discussed in \Sec{subsec:merge}. }\label{summary}
\end{figure}

In the following, we overview several recent results that pertain to DW--ZF interactions and were derived from the improved WKE.  In particular, the role of the ZF ``curvature'' $U''\doteq\pd_x^2 U$ will be emphasized. Sections \ref{subsec:traj} and \ref{subsec:KHI} are mostly based on published results. However, the discussion about pseudo-Hermiticity in \Sec{subsec:KHI}, the WKE-based derivation of the equations of state in \Sec{subsec:EOS}, and the discussion about merging and splitting of ZFs in \Sec{subsec:merge} are presented here for the first time. These results are based on the mHME and are summarized by a ``phase diagram'' in \Fig{summary}. Applications beyond the mHME will be discussed in \Sec{sec:curvature}.
\subsection{Drifton phase-space dynamics}
\label{subsec:traj}
\begin{figure}
\includegraphics[width=1\columnwidth]{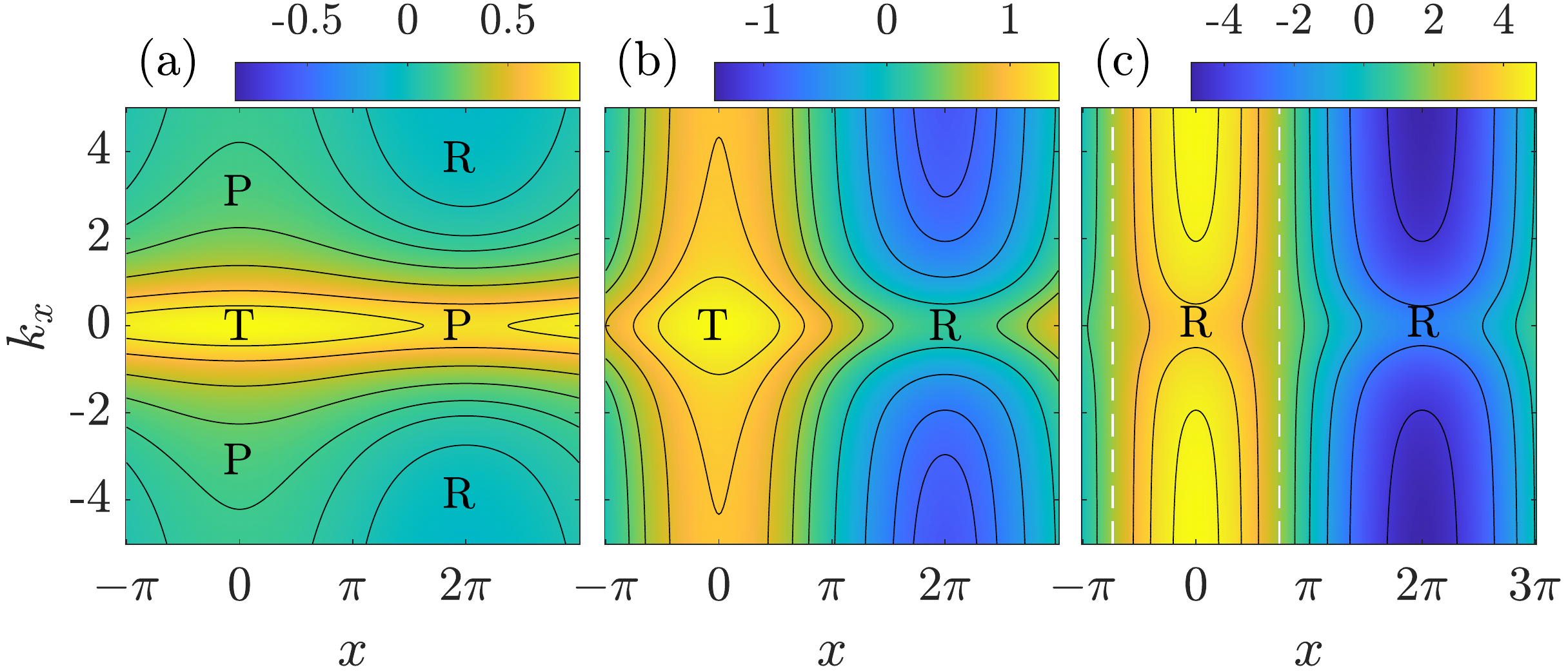}\caption{Contour plots of $\mc{H}$ [\Eq{eq:IWKE_H}]  with $U$ given by \Eq{eq:traj_U}. The parameters are $\kappa=1$ and $k_y=q=0.5$, so $u_{c,1} = 0.44$ and $u_{c,2} = 4$ [\Eq{eq:traj_uc}]. The subfigures correspond to three different regimes: (a)~$u = 0.2$, so $u < u_{c, 1}$; (b)~$u = 2$, so $u_{c, 1} < u < u_{c, 2}$; and (c)~$u = 10$, so $u > u_{c, 2}$.  The labels P, T, and R denote ``passing'', ``trapped'', and ``runaway'' trajectories. The white vertical dashed lines in (c) denote the locations where $U'' + \kappa=0$. }\label{traj}
\end{figure}

Like the traditional WKE, the improved WKE leads to ray equations of driftons:
\begin{equation}
\frac{d\bd{x}}{dt}=\frac{\pd \mc{H}}{\pd\bd{k}},\quad\frac{d\bd{k}}{dt}=-\frac{\pd \mc{H}}{\pd\bd{x}}.
\end{equation}
In stationary ZFs, where $d_t\mc{H}=\pd_t\mc{H}=0$, driftons move along constant-$\mc{H}$ contours in the $(x,k_x)$ phase space while $k_y$ is conserved. For simplicity, let us adopt the following ansatz for the ZF velocity:
\begin{equation}
U=u\cos qx.\label{eq:traj_U}
\end{equation}
Then, three different regimes are possible depending on how $u$ relates to the critical velocities\cite{Zhu18b,Zhu18c}
\begin{equation}
u_{c,1}\doteq\frac{\kappa}{2(1+k_y^2)-q^2},\quad u_{c,2}\doteq\frac{\kappa}{q^2}.\label{eq:traj_uc}
\end{equation}
The GO limit [\Eq{eq:WKE_GO}] corresponds to $u_{c,1} \ll u_{c,2}$. The corresponding topologies of the drifton phase space are illustrated in \Fig{traj} and differ in the presence of ``passing'', ``trapped'', and ``runaway'' trajectories. Passing trajectories vanish at $u > u_{c,1}$, and trapped trajectories vanish at $u > u_{c,2}$. Runaways are present at any $u$ and provide an important dissipation mechanism of DWs. Namely, driftons on runaway trajectories move towards $|k_x| \to \infty$ and are eventually damped by dissipation. This can be understood as shearing out of DW eddies by ZFs.

The first critical ZF amplitude $u_{c,1}$ can be considered as the boundary between ``weak'' and ``strong'' ZFs in the following sense. Consider the interactions between a ZF and a radially extended monochromatic DW with $k_x=0$ but $k_y\neq 0$. At $u<u_{c,1}$ most driftons reside on passing and trapped trajectories. However, passing trajectories vanish at $u>u_{c,1}$, when many driftons become runaways and transfer their energies to the ZF. That is when the ZF becomes efficient in shearing out  DW eddies. This transition from weak to strong ZFs has been studied in detail for the nonlinear stage of the modulational instability.\cite{Zhu19} There, the maximum ZF velocity is estimated as $u_{\max}\sim\gamma_{\rm MI}/k_y$ where $\gamma_{\rm MI}$ is the modulational-instability growth rate. Then, by comparing $u_{\rm max}$ with $u_{c,1}$, we find that the transition occurs at $\gamma_{\rm MI}/\Omega_{\bd{k}}\sim1$. This explains the numerically observed transition from oscillations to saturation of the ZF amplitude.\cite{Manfredi01,Connaughton10,Gallagher12}

The second critical ZF amplitude $u_{c,2}$ is attained when the ZF curvature $U''=-q^2 U$ satisfies (assuming $\kappa>0$)
\begin{gather}
\min (U''+\kappa) \leq 0.\label{eq:KHI_RK}
\end{gather}
This is known as the Rayleigh--Kuo criterion,\cite{Kuo49} which is relevant to the KHI and will be discussed in the next subsection. Note that the Rayleigh--Kuo criterion is missed in the traditional WKE, because the latter neglects $U''$, which effectively corresponds to  $u_{c,2} = \infty$.
\subsection{Kelvin--Helmholtz instability}
\label{subsec:KHI}
\begin{figure}
\includegraphics[width=1\columnwidth]{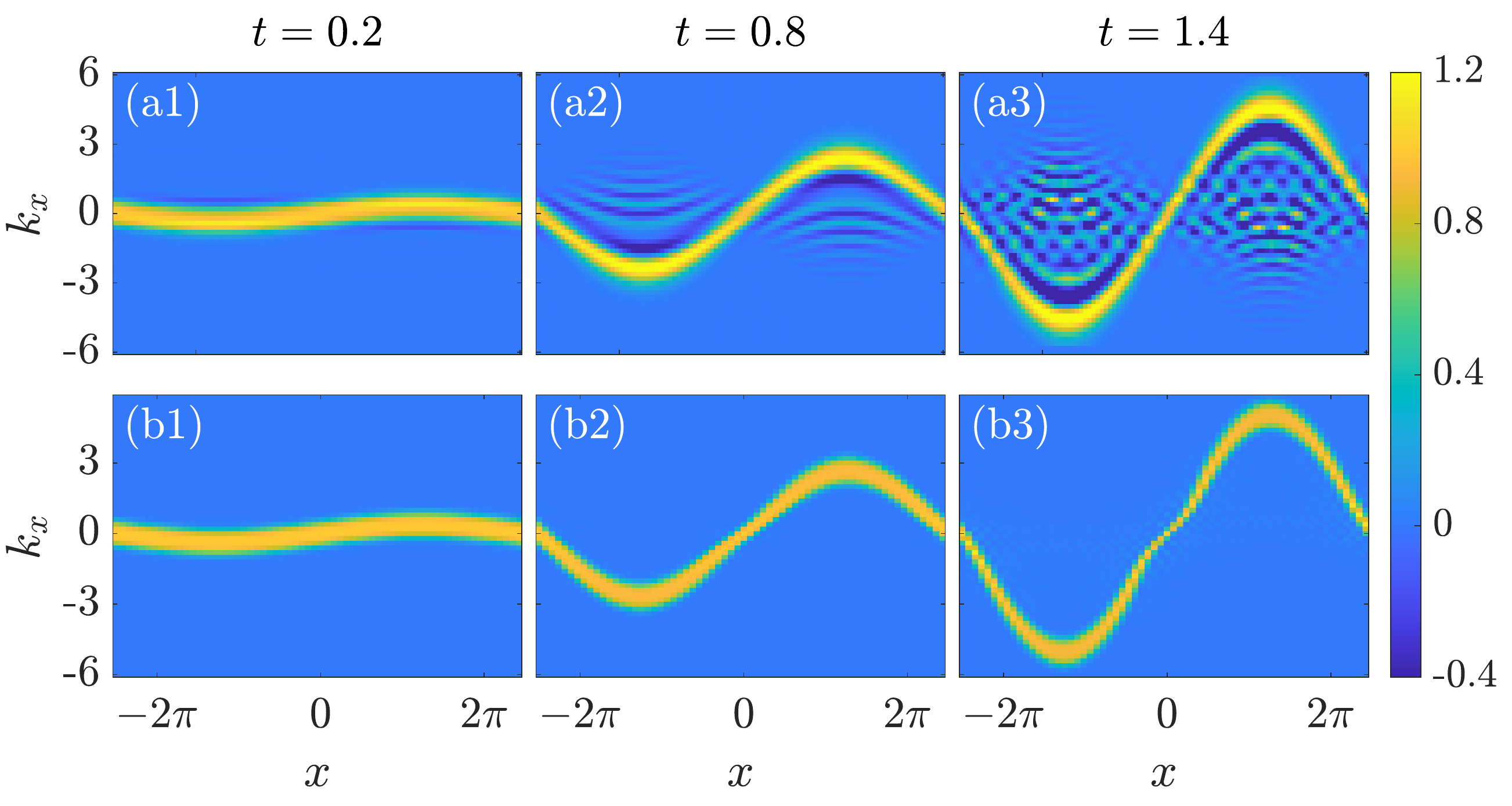}\caption{Numerical solutions of the Wigner--Moyal equation (upper row) and the improved WKE (lower row). The  numerical methods are described in Refs.~\onlinecite{Ruiz16,Parker18}. The initial conditions are $U=u\cos q x$ and $W=W_0 e^{-k_x^2/2\sigma^2}$, with $u=10$, $q=0.4$, $k_y=1$, $\kappa=1$, $\sigma=0.5$, and $W_0\ll 1$. Shown are $W/W_0$ at three different moments of time. The WKE solutions adequately approximate the exact  DW dynamics. The Rayleigh--Kuo criterion, $q^2u>\kappa$, is satisfied, but $q^2<1+k_y^2$. As a result, the drifton generation rate is superseded by the drifton dissipation at large $|k_x|$, so the KHI does not develop.}\label{KHI}
\end{figure}

The Rayleigh--Kuo criterion \eqref{eq:KHI_RK} is a generalization of the famous Rayleigh criterion in hydrodynamics\cite{Kuo49} and can be understood as follows. Let us consider the KHI as a Floquet mode $\tilde{w}(\bd{x}, t) = {\rm Re}[\tilde{w}(x)e^{ik_yy-i \omega t}]$ of \Eq{eq:IWKE_DW} with periodic $U$. Let $\hat{Q} \doteq \kappa + U''$, one can show that
\begin{gather}
\hat{H}\hat{Q} = \hat{Q}\hat{H}^\dag.\label{eq:KHI_SH}
\end{gather}
This makes $\hat{H}$ pseudo-Hermitian.\cite{Brizard94,Mostafazadeh02}  Consider a variable transformation $\tilde{w} = \hat{Q}^{1/2}\tilde{\eta}$, so \Eq{eq:IWKE_DW} becomes
\begin{gather}
i\partial_t \tilde{\eta} = 
\hat{\mathsf{H}} \tilde{\eta},
\quad
\hat{\mathsf{H}} \doteq \hat{Q}^{-1/2}
\hat{H}\hat{Q}^{1/2}.
\end{gather}
As long as $\hat{Q}$ is positive-definite, $\hat{Q}^{\pm 1/2}$ are Hermitian. Then, from \Eq{eq:KHI_SH}, the new Hamiltonian $\hat{\mathsf{H}}$ is Hermitian too, and thus its eigenvalues are real, \ie the corresponding Floquet modes are stable. This means that the instability requires that $\hat{Q}$ be \textit{not} positive-definite, and that is precisely the Rayleigh--Kuo criterion \eqref{eq:KHI_RK}. Using the common terminology,\cite{Mostafazadeh02} one can thereby consider the KHI as pseudo-Hermiticity breaking (also referred to as the parity--time symmetry breaking\cite{Qin19}). 

The Rayleigh--Kuo criterion  is attained at $u>u_{c,2}$, when all the driftons reside on runaway trajectories and eventually dissipate at large $|k_x|$. Then, the KHI occurs only when the drifton generation rate $\Gamma\propto q^3$ [\Eq{eq:IWKE_H}] exceeds the drifton dissipation rate.  A rigorous stability analysis in Ref.~\onlinecite{Zhu18a} shows that for a given $k_y$, a Floquet mode is stable if  $q^2\leq 1 + k_y^2$, and the unity originates from the assumed electron adiabatic response in the mHME. (The term ``tertiary instability'' was used in Ref.~\onlinecite{Zhu18a}, but what is actually discussed there is the KHI.) This means that no modes are unstable if $q^2 \leq 1$, as illustrated by numerical results in \Fig{KHI}. Therefore, in addition to the Rayleigh--Kuo criterion, the KHI  also requires
\begin{gather}
q ^2> 1.\label{eq:KHI_q}
\end{gather}
At $q^2 \leq 1$, the KHI is suppressed and the tertiary instability is more relevant, which is discussed in \Sec{sec:curvature}.
\subsection{Equation of state and solitary structures}
\label{subsec:EOS}
The WKE allows one to derive an ``equation of state'', which connects the drifton density with the ZF velocity. This is done as follows. Let us rewrite \Eq{eq:IWKE_W} as
\begin{multline}
\frac{\pd{W}}{\pd t}=\frac{\pd }{\pd k_x}\left[k_y\left(\frac{U'''}{\bar{k}^2}+U'\right)W\right]\\
+\frac{2k_xk_y(\kappa+U'')}{\bar{k}^4}\frac{\pd W}{\pd x}+F-{D},\label{eq:EOS_W}
\end{multline}
where we added forcing and dissipation for generality. Let us define the drifton density (DW enstrophy density)
\begin{equation}
Z(x,t)\doteq\frac{1}{2}\int\frac{d\bd{k}}{(2\pi)^2}{W}
\end{equation}
and  the drifton flux (DW enstrophy flux)
\begin{equation*}
J(x,t)=-\kappa\int\frac{d\bd{k}}{(2\pi)^2}\frac{k_xk_y}{\bar{k}^4}{W}.
\end{equation*}
Note that the true drifton flux is modified by $U''$, but the difference is small if $|U''|\ll\kappa$. Then, by integrating \Eq{eq:EOS_W} over $\bd{k}$, one obtains
\begin{equation}
\pd_t Z+(1+\kappa^{-1}U'')\pd_{x} J= S-\mu_{\rm DW} Z.\label{eq:EOS_Z}
\end{equation}
Here,  $S\doteq\int d\bd{k}\,F/8\pi^2$
is the drifton source and $\mu_{\rm DW}\doteq Z^{-1}\int d\bd{k}\,D/8\pi^2$
is the average drifton damping rate. Meanwhile, \Eq{eq:models_ZF} can be rewritten as
\begin{equation}
\pd_t U=-\kappa^{-1}\pd_{x}J-\mu_{\rm ZF} U,\label{eq:EOS_U}
\end{equation}
where we assumed a linear friction of ZFs, $\int dx\avg{d}=\mu_{\rm ZF}U$. Using the notation $ \delta Z\doteq Z-S/\mu_{\rm DW}$, one can rewrite Eqs.~\eq{eq:EOS_Z} and \eq{eq:EOS_U} as follows:
\begin{gather}
(\kappa+U'')(\pd_t+\mu_{\rm ZF})U=(\pd_t+\mu_{\rm DW})\delta Z.
\end{gather}
If $|U''| \ll \kappa$ (as we have already assumed) and $\pd_t \ll \mu$, which corresponds to a quasistatic case, one obtains
\begin{equation}
U\approx \kappa^{-1} (\mu_{\rm DW}/\mu_{\rm ZF})\delta Z.\label{eq:EOS_EOS1}
\end{equation}
However, for conservative systems  ($F=D= 0$), one has
\begin{equation}
U\approx \kappa^{-1}\delta Z+c,\label{eq:EOS_EOS2}
\end{equation}
where $c$ is some integration constant.
Note that for the conservative system,  $S/\mu_{\rm DW}$ can still remain finite and can be considered as the initial condition for $Z$.

The equation of state, originally reported in Ref.~\onlinecite{Zhou19} (and recently derived using a different approach in Ref.~\onlinecite{Krasheninnikov21}), has been used to study the nonlinear stage of the modulational instability of quasmonochromatic DWs. For a DW of the form $\tilde{w}={\rm Re}(\psi e^{i\bd{k}\cdot\bd{x}-i\Omega_{\bd{k}}t})$ with slowly varying envelope $\psi$, \Eq{eq:EOS_EOS2} gives $U=|\psi|^2/4\kappa+c$. Then, \Eq{eq:IWKE_DW} leads to the following nonlinear Schr\"odinger equation:
\begin{equation}
i(\pd_t+v_g\pd_x)\psi=-(\chi/2)\pd_x^2\psi+k_y|\psi|^2\psi/4\kappa,\label{eq:EOS_NLSE}
\end{equation}
where $v_g\doteq\pd\Omega_{\bd{k}}/\pd k_x$ is the same group velocity as before and $\chi\doteq\pd^2\Omega_{\bd{k}}/\pd k_x^2=2\kappa k_y(4k_x^2-\bar{k}^2)/\bar{k}^6$.  Also, $c$ has been absorbed into the phase of $\psi$. Equation \eqref{eq:EOS_NLSE} supports soliton solutions of the form
\begin{gather}
\psi=2\eta\sqrt{-\frac{\kappa\chi}{k_y}}\frac{e^{i\chi\eta^2 t/2}}{\cosh[\eta(x-v_gt)]},
\end{gather}
with a free parameter $\eta$. Similar solitary structures are commonly produced upon saturation of the modulational instability.\cite{Zhou19,Zhou20}  Equation \eqref{eq:EOS_NLSE} was also derived in the past using alternative approaches,\cite{Champeaux01,Dewar07book} but the one presented here is, arguably, more transparent. In particular, the equation of state indicates a correlation between soliton formation and drifton trapping at peaks of $U$ (\Fig{traj}).
\subsection{Splitting and merging of zonal flows}
\label{subsec:merge}
\begin{figure}
\includegraphics[width=1\columnwidth]{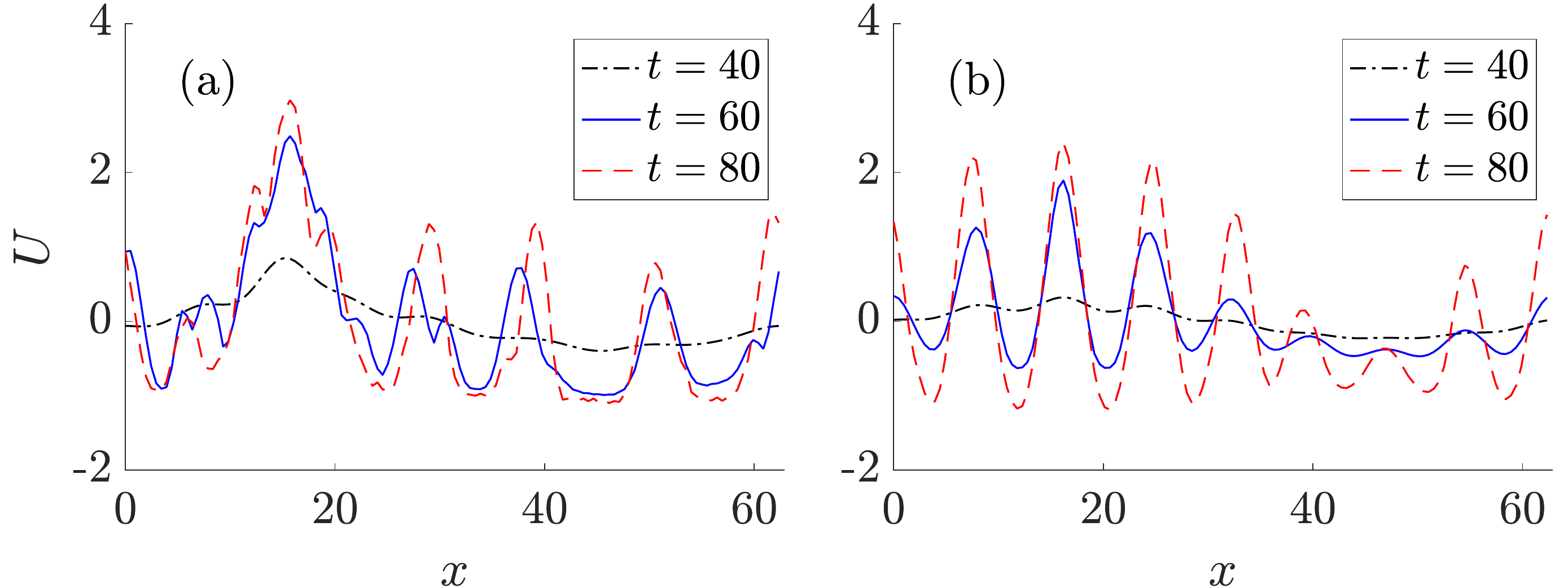}\caption{Results from numerical simulations of (a) the mHME and (b) the improved WKE.  The initial conditions are $\tilde{w}=w_0\cos p y$, $U=u\sin qx$, and small random noise is added. This corresponds to $W=\pi^2 w_0^2\delta(k_x)\delta(|k_y|- p)$ as the initial condition of the WKE. No forcing and dissipation are included except a small fourth-order hyperviscosity $d=10^{-4}\nabla^4 w$. The parameters are $\kappa=1$, $p=0.5$, $q=0.1$, $w_0=2$, and $u=0.1$. Shown are the ZF velocity at three different moments of time. The initially large-scale ZFs split into smaller-scale ZFs.}\label{split}
\end{figure}

Unlike the traditional WKE, the improved WKE predicts that the maximum of $\gamma_{\rm ZI}$ corresponds to a finite $q$, where $\gamma_{\rm ZI}\doteq{\rm Im}\,\omega$ is the zonostrophic-instability growth rate [\Eq{eq:IWKE_dispersion}]. Even more importantly, the improved WKE can also be used to estimate the characteristic scale of naturally formed ZFs in the saturated state.  Although there is no comprehensive theory yet that would unambiguously predict the ZF wavenumber, one can expect that if the initial conditions are such that ZFs are produced with a ``nonoptimal'' $q$, the ZFs  will exhibit splitting or merging  until the optimal scale is reached.   

The splitting of ZFs can be easily understood. Large-scale ZFs form a nearly homogeneous background, so the zonostrophic instability can develop at smaller scales just like in homogeneous plasma, which can be interpreted as ZF splitting. This process is illustrated in \Fig{split}.

\begin{figure}
\includegraphics[width=1\columnwidth]{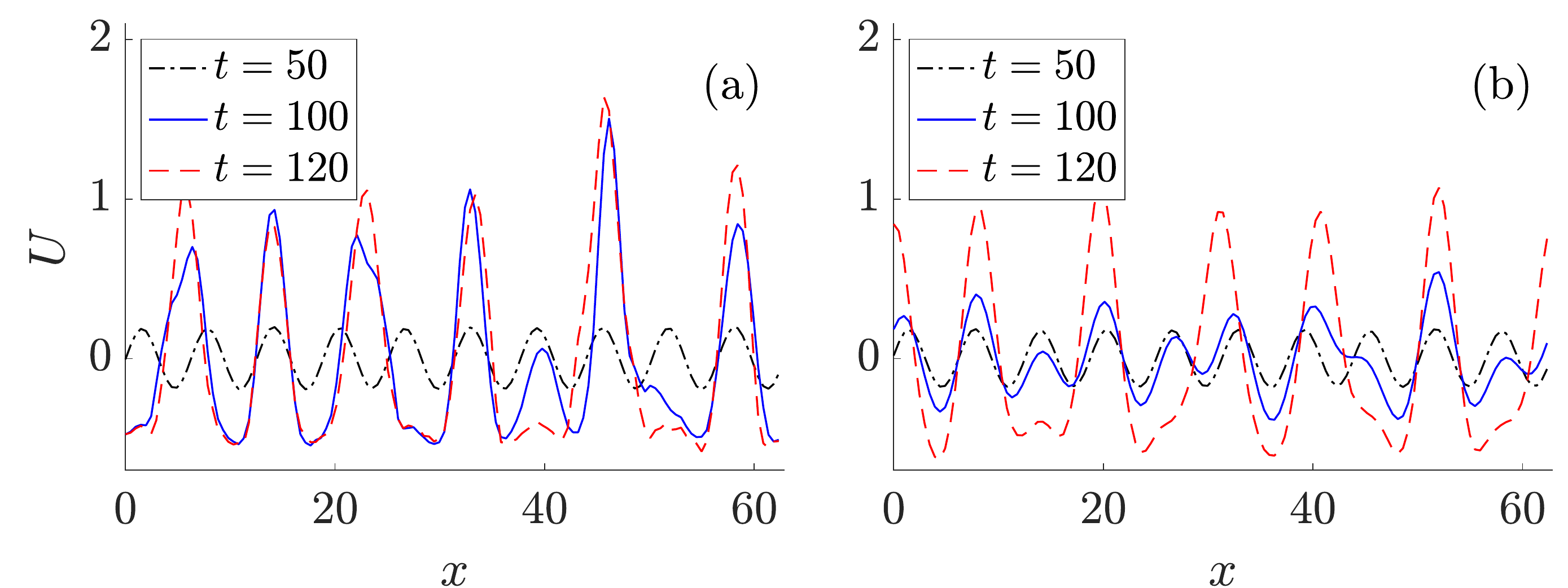}\caption{Same as \Fig{split} but with $w_0=1.4$ and $q=1$. The initially small-scale ZFs merge into larger-scale ZFs.}\label{merge}
\end{figure}

Merging of small-scale ZFs is also readily seen in simulations (\Fig{merge}) but is less intuitive. To understand it qualitatively, let us consider the drifton Hamiltonian given by \Eq{eq:IWKE_H}. Since the initial conditions in \Fig{merge} correspond to $k_y=\pm p$ and $k_x=0$, we assume that $k_x^2\ll k_y^2$  so that $\bar{k}^{-2}$ can be expanded in $k_x^2$.  Let us also adopt the equation of state \eqref{eq:EOS_EOS2}. Then, the Hamiltonian can be approximated as
\begin{equation}
\mc{H}\approx-\frac{k_y}{\kappa}\left(\frac{k_x^2}{2m}+V\right)+\mc{H}_0,
\end{equation}
where $\mc{H}_0=\kappa k_y/(1+k_y^2)$, $ m\doteq(1+k_y^2)^2/2\kappa^2$, and
\begin{gather}
V=\left(\frac{q^2}{1+k_y^2}-1\right)\delta Z.\label{eq:merge_V}
\end{gather}
It is seen then that a drifton behaves like a classical particle with mass $m>0$, momentum $k_x$, and potential energy $V$. (The overall coefficient $-k_y/\kappa$ can be eliminated by redefining the time variable and hence is unimportant.) Because $\delta Z$ is the drifton-density perturbation, \Eq{eq:merge_V} suggests that states with $q^2 > 1+ k_y^2$ are energetically unfavorable. Hence, one can expect that the system tends to reduce $q$, corresponding to ZF merging. This argument is only qualitative, because the applicability of the WKE at large $q$ is questionable. Still, it suggests the following upper bound on $q$ for stable ZFs:
\begin{gather}
q_{\rm max} \sim \sqrt{1 + k_y^2},
\end{gather}
which is consistent with the KHI analysis in \Sec{subsec:KHI}.

Note that the above arguments assume negligible forcing and dissipation. In the opposite limit, when forcing and dissipation are strong, the ZF amplitude has been shown to obey the Ginzburg--Landau equation near the threshold of the zonostrophic instability.\cite{Parker13,Parker14,Bakas19} Then, ZFs with too large or too small $q$ are subject to the Eckhaus instability, so the scalings may be different. 
\section{Role of the zonal-flow curvature  beyond the modified Hasegawa--Mima model}
\label{sec:curvature}
Although the mHME model captures many important aspects of DW--ZF interactions, it is fundamentally limited in that it supports no primary instabilities. To mimic a primary instability, the traditional WKE often includes a forcing  $F=\gamma_{\bd{k}}N$,
where $\gamma_{\bd{k}}$ is positive in a certain range of $\bd{k}$.\cite{Kaw01,Malkov01PP,Singh14,Wang09,Sasaki18} However, it does not properly capture the dependence of $\gamma_{\bd{k}}$ on the ZF curvature, which can be crucial. To understand this effect, we consider the mTHE model [\Eq{eq:models_TH}] with a built-in primary DW instability.  In this case, the improved WKE has the same form as the one for the mHME [\Eq{eq:IWKE_W}] but $\bar{k}^2=1+k^2-i\delta_{\bd{k}}$, where $\delta_{\bd{k}}$ gives rise to the primary instability. Then, to the lowest order the $U'''$ term may be neglected, but $U''$ generally should be retained:\cite{Zhu20JPP}
\begin{gather}
\label{eq:curvature_H}
\mc{H}=k_yU+{\rm Re}\,\Omega_{\bd{k}}(1+U''/\kappa),\\
\Gamma={\rm Im}\,\Omega_{\bd{k}}(1+U''/\kappa).
\end{gather}
It is seen that $U''$ directly modifies the DW frequencies and growth rates. Even though  $q$ is assumed small, the WKE does not pose a restriction on the ZF amplitude; hence, $|U''|\sim q^2 |U|$ can be comparable to $\kappa$, in which case the DW frequencies and growth rates are substantially modified. Importantly, this modification is inhomogeneous in space;  the growth rate is reduced at ZF peaks ($U''<0$) and is enhanced at ZF troughs ($U''>0$).  (These conclusions also extend to more complex models of DW turbulence\cite{Zhu20PRL,Ivanov20,Hallenbert20} and have also been verified independently through deep learning.\cite{Heinonen20})

The dependence of $\mc{H}$ and $\Gamma$ on $U''$ leads to the tertiary instability of ZFs.\cite{Rogers00} In particular, $\mc{H}$ describes the tertiary-mode structures while $\Gamma$ describes their growth rates.\cite{Zhu20PRL,Zhu20JPP} (Since $\Gamma\propto{\rm Im}\,\Omega_{\bd{k}}$ vanishes at $\delta_{\bd{k}}=0$, the tertiary instability is missed in the mHME.) Since the tertiary instability determines whether ZFs can suppress DWs, the threshold of DW turbulence onset is also $U''$-dependent. This has led to an explicit estimate of the Dimits shift within the mTHE.\cite{Zhu20PRL,Zhu20JPP} Also, as discussed in Ref.~\onlinecite{Zhu20JPP}, the improved WKE can only provide a qualitative description of the tertiary instability,  while achieving quantitative agreement requires application of the complete Wigner--Moyal approach. 

The tertiary instability is often confused with the KHI,\cite{Kim03,Numata07,Li18,Zhu18a} but the recent studies reveal that their physical mechanisms are very different.\cite{Zhu20JPP,Zhu20PRL,Hallenbert20} The tertiary instability extracts energy from the background gradients just like the primary instability, while the KHI extracts energy from the flow shear. In fact, the tertiary instability can be considered as a primary instability modified by $U''$. Since ZFs usually form large-scale structures, the tertiary instability is more relevant than the KHI, because the latter is stabilized at such scales (assuming that the electron response is close to adiabatic; see \Sec{subsec:KHI}).  Similar conclusions also apply within the modified Hasegawa--Wakatani model \cite{Zhu20PRL} and within the ITG model from Ref.~\onlinecite{Rogers00}.\footnote{See Supplemental Material of Ref.~\onlinecite{Zhu20PRL}} 
\section{Summary}
\label{sec:summary}
In this paper, we overview the recent advances in the wave-kinetic theory of inhomogeneous DW-ZF interactions\cite{Parker16,Ruiz16,Ruiz19,Zhu18a,Zhu18b,Zhu18c,Zhu19,Zhu20JPP,Zhu20PRL,Zhou19,Zhou20,Tsiolis20}
and present a consolidated high-level physics picture of ZF physics, mostly based on the mHME model. A special focus  is made on understanding drifton phase-space dynamics, associated solitary structures, merging and splitting of ZFs,  as well as the KHI and the tertiary instability.  We also briefly mention the connection between these  findings and the recent progress in analytic understanding of the Dimits shift.
\begin{acknowledgments}
This work was supported by the US DOE through Contract No. DE-AC02-09CH11466. 
\end{acknowledgments}

\section*{Data Availability}
Digital data can be found in DataSpace of Princeton University.\footnote{See \url{http://arks.princeton.edu/ark:/88435/dsp01ws859j72v} for digital data.}

\bibliography{references}

\end{document}